# Valeur des tests PACE et CTB_ELISA dans le diagnostic de la peste porcine classique (PPC) et le contrôle de qualité du vaccin correspondant à Madagascar

O. F. MAMINIAINA[(1)], KOKO[(2)], J. J. RAJAONARISON[(2)] RAZAFINDRAKOTO[(3)], J. RAVAOMANANA [(1)] et A. D. SHANNON [(4)]

[(1)] *Laboratoire de Diagnostic du Département de Recherche Zootechnique et Vétérinaires / Centre National de la Recherche Appliquée au Développement Rural (DRZV/ FOFIFA). Ministère de l'Education National et de la Recherche Scientifique, Antananarivo Madagascar*
[(2)] *Faculté de Médecine – Filière Vétérinaire). Ministère de l'Education National et de la Recherche Scientifique, Antananarivo Madagascar*
[(3)] *Laboratoire de Virologie de l'Institut Malgache des Vaccins Vétérinaires (I.M.VA.VET.) Ministère de l'Education National et de la Recherche Scientifique, Antananarivo Madagascar*
[(3)] *Elizabeth Macarthur Agricultural Institute, NSW Agriculture, Australia*

## Résumé :

A partir de 1994, nous avons commencé à utiliser l'ELISA (Enzyme Linked Immunosorbent Assay) dans le diagnostic de la PPC. Il s'agit d'un ELISA de capture d'antigènes (PACE) éventuellement contenus dans les prélèvements. L'avantage de ce test vient du fait qu'il est totalement indépendant des cultures cellulaires. De plus, il est rapide : le résultat peut être obtenu en moins de 36 heures. Une étude de sa standardisation effectuée en Australie a donné une sensibilité (Se) de 99%, une spécificité (Sp) voisine de 100 % et une valeur prédictive négative (VPN) de 99,7 %. De par sa haute spécificité, le test est donne un résultat négatif à tous les vrais négatifs, autrement dit, aux négatifs du test correspondent les vrais négatifs. Une variante de l'ELISA de capture, le CTB-ELISA ou complex trapping blocking ELISA permet de doser la quantité d'anticorps dirigés contre la protéine non structurale, la p80 (ou $NS_3$), contenus dans les sérums des animaux. L'évaluation du taux d'anticorps anti-$NS_3$ constitue une excellente appréciation du taux d'anticorps neutralisants car le coefficient de corrélation entre ces deux types d'anticorps, le premier obtenu par CTB-ELISA, et le second par la séroneutralisation (VNT), est très élevé (r= 0,98).

Les deux tests étant capables, l'un de détecter des antigènes pestiviraux l'autre de mesurer des anticorps spécifiques de chacun des groupes, constitue un excellent outil pour le contrôle qualitatif de vaccin anti-PPC.

**Mot clés:** *Peste porcine classique- Pestivirus - Diagnostic – Contrôle de qualité - Madagascar*

# Value of two ELISA tests in the diagnosis of classical swine fever (CSF) and quality control of the corresponding vaccine in Madagascar

### Abstract

At the central livestock laboratory, the diagnostic of CSF was performed by using the exaltation of Newcastle disease virus test (END) on swine testis cells. Although it is highly specific, this test has the disadvantage to be time consuming. Besides all problems related to cell culture, the result is obtained within about 10 days. However, one of the qualities of the diagnostic method is its rapidity because the decision depends on it. This is why the direct immunofluorescence (DIF) technique has taken over the END test since 1984. The DIF, applied either on spleen smear or on sensitive cells previously infected with tissue extract is quick since the result can be achieved within two hours (smear) or 72 hours (cell culture inoculation). Nevertheless, it requires good technical skill and the use of microscope of good quality to correctly detect the viral antigen and also to select specific fluorescence from non-specific one.

Since 1994, we have been running the ELISA (Enzyme Linked Immunosorbent Assay) in the diagnostic of CSF. It concerns an antigen capture ELISA for pestivirus (PACE) eventually included in samples. This test does not require cell culture at all, that is its great advantage. In addition, it is rapid: the result can be obtained in less than 36 hours. Study of its standardization in Australia demonstrated a sensivity (Se) of 99%, a specificity (Sp) around 100% and a negative predictive value (NPV) reaching 99,7%. Thanks to its high specificity, the test is theorically able to score negative all the true negatives, in other words, test negatives correspond to true negatives.

A variant of capture ELISA, the complex trapping blocking ELISA (CTB-ELISA) allows to assay the quality of serum antibodies against p80, a non structural protein (NS3). Anti-NS3 serum antibody evaluation constitutes an excellent appreciation of neutralizing antibody level because the correlation rate between this two types of antibodies, the first one measured by CTB-ELISA and the second one by virus neutralization test (VNT) is very high (r = 0,98).

Both tests being capable either to detect pestiviral antigens or to measure specific antibodies of each group, represent a very good tool for quality control of vaccines against CSFV.

I. **INTRODUCTION**

La peste porcine classique ou PPC est une maladie inoculable, contagieuse, spécifique aux Suidés (porc domestique et sanglier), causée par un virus à ARN positif, appartenant au genre *Pestivirus* [10] (Westaway et al., 1985). Ce groupe comprend outre le virus de la PPC ou « *classical swine fever virus* » (CSFV) le virus de la maladie des muqueuses des bovins (BDVD : *Bovine viral Diarrhea Virus*) et celui de la maladie de Border chez les moutons (BDV : *Border Disease Virus*), qui sont tous antigéniquement apparentés [3]. La PPC cause une perte importante dans l'élevage porcin malgache avec la maladie de Teschen, depuis 1965, où elle était signalée pour la première fois dans le pays. Quatre techniques ont été utilisées au laboratoire central de l'Elevage :

\* soit pour le diagnostic (laboratoire de diagnostic): test END ou *Exaltation of Newcastle Disease* sur cellules testiculaires d'un porcelet (TP) [4, 5], reproduction de la maladie sur animal sain (1965–1984) et test immunofluorescence directe ou IFD (1984 – 1994) [1].

\* soit pour le contrôle du vaccin suipestique (Unité de Production des vaccins): test d'innocuité sur porc couplé à une épreuve virulente (1965 – 1997).

Ces différentes techniques bien que très spécifiques présentent des inconvénients :

- lourdes à réaliser pour le diagnostic de routine (END, épreuve virulente, séroneutralisation)
- nécessitent un microscope à fluorescence de qualité et une bonne expérience professionnelle (IF)
- risque de dissémination de virus sauvage (épreuve virulente)

Actuellement, grâce au projet AIEA/FAO ou Agence internationale de l'énergie atomique / *Food and Agriculture organization*, les deux laboratoires disposent de deux tests ELISA (*Enzyme Linked Immunosorbent Assay*) qu'ils utilisent dans le diagnostic de la PPC (*Pestivirus Antigen Capture ELISA* ou PACE) [8], le contrôle de qualité du vaccin correspondant et la recherche des contaminants Pestiviraux (PACE et *Complex Trapping Blocking ELISA* ou CTB-ELISA) [9].

## II. MATERIEL ET METHODES

### 1. Matériel

#### a. *Anticorps monoclonaux (mAbs)*

Pour la réalisation de ces deux tests, 4 groupes d'anticorps monoclonaux (mAbs) de souris BALB/C produits par le laboratoire EMAI-NSW ou Elisabeth Macarthur Agricultural Institute – New South Wales (Australia) ont été utilisés à la dilution de 1/200. Il s'agit du :

* R 882 : mélange de 3 clones $N_2B_{12}$, $2NA_1$ et $INB_5$ (v/v/v) : mAbs spécifiques de la protéine structurale d'enveloppe $E_2$ du BVDV [8, 9].

* R 1048 : mélange de 3 clones $P_3C_{12}$, $2NB_{12}$ et $P_4G_{11}$ (v/v/v) : mAbs spécifiques de la protéine non structurale p80 ou $NS_3$, antigène de groupe qu'on retrouve dans toutes les cellules infectées par les *Pestivirus* (BVDV, CSFV et BDV) [8, 9]

* Clone Tü 24/10 : mAbs spécifiques de la protéine structurale d'enveloppe $E_2$ du virus CSFV [4].

* R 769 : clone 14/4C4 : mAbs négatifs. Ce sont des anticorps monoclonaux spécifiques d'une protéine structurale de *Lentivirus* des caprins qui ne présentent aucune réaction croisée avec les membres du *Pestivirus* [8, 9].

Il est à remarquer que :

- aucun groupe des mAbs ne reconnaît un pestivirus autre que celui qui lui est spécifique (absence de réactions croisées)
- chaque groupe des mAbs reconnaît à 100 % le virus qui lui est spécifique
- les mAbs négatifs ne reconnaissent aucun des *Pestivirus*

#### b. *Anticorps polyclonaux (pAbs)*

L'anticorps polyclonal (pAbs) utilisé est un antisérum (IgG) de chèvre, reconnaissant toutes les souches des *Pestivirus* (BVDV, CSFV et BDV). Ces pAbs ont été produits selon la méthode de Meyling (1984), en utilisant 7 souches NSW de *Pestivirus* [8, 9], en provenance de EMAI (Elisabeth Macarthur Agricultural Institute – Australia). La dilution d'emploi a été de 1/2000.

c. *Antigènes* :

* Antigènes BVDV positifs R 1237 U, R 624 et R 710.

* Antigènes CSFV positifs R 1327-U, R 838-U et R 883-U.

* Antigènes de contrôle négatif : R 837-U, R 930.8.

Tous ces réactifs ont été préparés par le laboratoire de Santé Animale d'Australie (AAHL ou Australian Animal Health Laboratory).

d. *Sérums*

   i. *Sérums bovins*
   - 13 lots de sérum destinés à la production cellulaire, récoltés sur des bovins abattus à l'abattoir frigorifique d'Antananarivo entre 1994 et 1998. Chaque lot est composé de 10 à 13 sérums prélevés sur 10 à 13 bovins.
   - 8 sérums de veau provenant de la ferme école BEVALALA, à 8 km au Sud d'Antananarivo.
   - 18 sérums de veau de 3 à 8 mois d'âge, prélevés dans la ferme d'Etat de KIANJASOA (Moyen-Ouest malgache).
   - R 947 : sérum de bovin contrôlé positif en Pestivirus fourni par l'EMAI (C1 : 1/100 ; C2 :1/250 ; C3 :1/500)
   - NHS : sérum normal de cheval sans Pestivirus provenant de l'Australian Animal Health Laboratory (AAHL).
   - ABS : sérum normal de bovin adulte produit par l'EMAI

   ii. *Sérums porcins*
   - 30 sérums de porcs vaccinés avec RAMJIVAX ®, provenant d'une ferme privée appartenant à Mr RAZAFIMAHEFA, sise à Ambohitrarahaba, un village situé à 5 km au Nord de la Capitale
   - 40 sérums de porcs vaccinés avec RAMJIVAX ®, provenant de la ferme école BEVALALA

- 10 sérums de porcs immunisés le même vaccin et 5 non vaccinés de l'animalerie de l'Institut Malgache des Vaccins Vétérinaires (IMVAVET)
- Des lots de sérums de référence (négatif et positif) fournis par le Laboratoire de Santé Animale d'Australie (AAHL) et dépourvus de Pestivirus : sérum AAHL négatif, sérums AAHL positifs ((#327, #330 et #329)

e. *Leucocytes (PBL ou Peripheral Blood Leucocyte)*

  i. *PBL de porc*
  - Un échantillon de PBL d'un porc malade d'une ferme à SABOTSY NAMEHANA (élevage encadré par la Maison du Petit Elevage ou MPE, situé à environ 8 km au nord de la Capitale, Antananarivo), récolté lors de l'épidémie porcine de 1998.

  ii. *PBL de bovin*
  - 114 PBL, récoltés à partir de sang hépariné des bovins sacrifiés à l'abattoir d'Antananarivo en 1998.
  - 16 échantillons de PBL provenant de la ferme d'Etat de KIANJASOA.

f. *Echantillons d'organes*
  - 18 échantillons d'organes, prélevés sur des porcs malades parvenus au laboratoire de Diagnostic (DRZV/FOFIFA) pendant l'épidémie porcine de 1998. Ces prélèvements portent des numéros de la série D (D 123 à D 316).
  - DRZV positif et DRZV négatif : antigènes préparés à partir de la rate des animaux provenant de l'animalerie du Département de Recherche Zootechnique et Vétérinaires expérimentalement infectés (DRZV pos) ou indemnes de CSFV (DRZV nég).

## 2. Méthodes

### a. *Préparation des extraits cellulaires*

L'extraction de virus du culot de leucocytes (PBL) ou tissus (rate, ganglion, intestin) se fait par la méthode préconisée par Shannon [7] dont l'essentiel est résumé comme suit :

#### i. *A partir des leucocytes du sang périphérique (PBL)*

- Centrifuger à +4°C à 3000 rpm pendant 20 min du sang hépariné dans un tube de 10ml
- Prélever les leucocytes formant une voile blanche qui se trouve entre le culot sanguin et le surnageant, à l'aide d'une pipette Pasteur plastique
- Placer les leucocytes dans un tube de 10 ml et y ajouter 5 ml de $NH_4Cl$ (Sigma) 0,17 M. Agiter le tube puis attendre 10 min
- Remplir le tube avec du tampon phosphate ou PBS et agiter
- Centrifuger pendant 10 min à 3000 rpm
- Jeter le surnageant et additionner le culot de 1 ml de PBS/1% NP-40 (Nonidet P-40, Sigma).
- Agiter avec un vortex toutes les 30 min pendant 2 heures à la température de laboratoire
- Centrifuger pendant 10 min à 3000 rpm à +4°C
- Utiliser le surnageant pour le test

#### ii. *A partir d'échantillons de tissus*

Les tissus doivent être placés à +4°C s'ils ne sont pas utilisés immédiatement.

- Prélever un ou deux morceaux d'environ 2g de rate, de ganglion (mésentérique ou inguinal), d'intestin et du poumon pour chaque animal testé. La rate et les ganglions constituent toutefois les organes de choix.
- Placer les tissus dans un tube de 10 ml et les découper à l'aide d'une paire de ciseaux.
- Ajouter 5 ml de PBS/1% NP-40.

- Agiter avec un vortex toutes les 10 min pendant 2 heures à la température de laboratoire.
- Centrifuger pendant 30 min à 3000 rpm.
- Utiliser le surnageant pour le test.

**b. *L'ELISA de capture ou Pestivirus Antigène Capture ELISA (PACE)***

Le principe du PACE est basé sur la capture de l'antigène pestiviral par un antisérum polyclonal de chèvre de haut titre et la détection de l'antigène capturé avec un mélange d'anticorps monoclonaux hautement spécifiques. L'amplification par addition d'un complexe avidine-biotine (ABC) augmente la sensibilité du test.

Les résultats sont obtenus en calculant le rapport ou ratio S/N (*Signal to Noise ratio*) selon la formule suivante :

$$\text{Ratio S/N} = DO_{mAbs\ positif} / DO_{mAbs\ négatif}$$

$DO_{mAbs\ positif}$ = Densité optique d'un échantillon en présence des mAbs BVDV (R 882), mAbs spécifiques du groupe (R 1048) ou mAbs CSFV (Clone Tü 24/10)

$DO_{mAbs\ négatif}$ = Densité optique d'un échantillon en présence des mAbs négatifs (R 769)

L'interprétation des résultats est résumée dans le tableau I

Tableau I : *Interprétation des résultats PACE*

| Résultats | Ratio S/N | $DO_{mAbs\ positif}$ |
|---|---|---|
| Négatif | ≤ 0,150 | ≤ 0,2 |
| Douteux | ]0,150 – 2,0[ | ]0,200 – 0,300[ |
| Positif | ≥ 2,0 | ≥ 0,300 |

**c. *L'ELISA de compétition ou Complex Trapping and Blocking Elisa (CTB-ELISA)***

Le principe de ce test est similaire à celui du PACE, avec cette différence que les anticorps du sérum à doser et les anticorps monoclonaux sont mis en compétition pour l'occupation des épitopes de l'antigène viral ($NS_3$ ou $E_2$)

Les résultats présentés sous forme de pourcentage d'inhibition (%INH), sont obtenus selon la formule suivante :

$$\%INH = ((DO_S/DO_{CN}) \times 100) - 100$$

$DO_S$ = Densité optique du sérum à doser

$DO_{CN}$ = Densité optique du contrôle négatif

L'interprétation des résultats est résumée dans le tableau II suivant :

Tableau II : *Interprétation des résultats du CTB-ELISA*

| % INH | Sérum |
|---|---|
| < 15 | Négatif |
| ]15 - 30] | Douteux |
| ]30 - 100] | Positif |

## III. RESULTATS ET INTERPRETATION

### A. Application du PACE pour la détection des antigènes pestiviraux

#### 1. Recherche d'antigènes $NS_3$ et/ou $E_2$ dans les leucocytes ou rates des porcs

Analysés par ce système ELISA PACE, les échantillons d'organes prélevés à l'occasion de cette épidémie porcine de 1998 à Madagascar ont donné majoritairement des résultats négatifs en présence des monoclonaux R 1048, spécifiques du groupe Pestivirus (Tab.III). Quatre prélèvements ont fait exception en donnant des résultats positifs, ce qui indique qu'ils renferment des antigènes du groupe Pestivirus. Par contre, ceux ayant servi au diagnostic de la PPA et qui proviennent des fermes encadrées par la MPE ont tous donné des résultats négatifs au PACE (Tab.III). Le test de confirmation effectué avec les monoclonaux spécifiques du virus de la maladie des muqueuses (R 882) a bien révélé qu'il ne s'agissait pas du BVDV (Tab.IV) et étant donné que le $3^e$ groupe de *Pestivirus* (le virus de la maladie de Border ou BDV) n'intéresse que les moutons, les antigènes de groupe détectés sur les 4 échantillons appartiennent par conséquent au virus CSFV. Les porcs ayant fourni ces échantillons ont dû en plus contracter la peste porcine classique. Les résultats des contrôles (positifs et négatifs) sont parfaitement cadrés dans les limites prévues et en conséquence garantissent la validité des tests effectués (Tab.III et Tab.IV).

Tableau III : Diagnostic des échantillons d'organes et PBL des porcs par PACE

| Origine | Echantillons | Nombre | Résultats densité optique (DO) mAbs R 1048 | mAbs R 769 | Interprétation S/N | Statuts |
|---|---|---|---|---|---|---|
| DRZV | D 123 | 1 | 0,179 | 0,166 | 1,0 | Négatif |
| | D 124 | 1 | 0,150 | 0,125 | 1,2 | Négatif |
| | D 125 | 1 | 0,764 | 0,175 | 4,4 | **Positif** |
| | D 169 | 1 | 0,637 | 0,199 | 3,2 | **Positif** |
| | D 170 | 1 | 0,484 | 0,174 | 2,8 | **Positif** |
| | D 174 | 1 | 0,130 | 0,113 | 1,2 | Négatif |
| | D 175 | 1 | 0,174 | 0,122 | 1,4 | Négatif |
| | D 239 | 1 | 0,057 | 0,059 | 0,9 | Négatif |
| | D 242 | 1 | 0,180 | 0,165 | 1,1 | Négatif |
| | D 243 | 1 | 0,099 | 0,100 | 0,9 | Négatif |
| | D 248 | 1 | 0,085 | 0,077 | 1,1 | Négatif |
| | D Bev | 1 | 0,119 | 0,114 | 1,0 | Négatif |
| | D 308 | 1 | 0,139 | 0,082 | 1,7 | Négatif |
| | D 311 | 1 | 0,167 | 0,117 | 1,1 | Négatif |
| | D 313 | 1 | 0,422 | 0,124 | 3,4 | **Positif** |
| | D 314 | 1 | 0,157 | 0,091 | 1,7 | Négatif |
| | D 315 | 1 | 0,153 | 0,085 | 1,8 | Négatif |
| | D 316 | 1 | 0,077 | 0,072 | 1,0 | Négatif |
| MPE | PBL* | 1 | 0,189 | 0,171 | 1,1 | Négatif |
| Contrôle (AAHL) | R 1327 U | 3 tests | 0,653 | 0,126 | 5,1 | Positif |
| | R 930.8 | 2 tests | 0,042 | 0,039 | 1,0 | Négatif |
| | R 837 U | 1 test | 0,176 | 0,112 | 1,5 | Négatif |
| | DRZV pos | 3 tests | 0,665 | 0,050 | 13,3 | Positif |
| Lab Diag | DRZV nég | 3 tests | 0,07 | 0,055 | 1,2 | Négative |

*:Leucocyte d'un porc malade pendant l'épizootie de PPA

Tableau IV : Détermination de l'origine des antigènes de groupe des 4 échantillons positifs PACE

| Origine | Echantillons | Nombres | Résultats densité optique (DO) mAbs R 882 | mAbs R 769 | Résultats et interprétation S/N | Statuts |
|---|---|---|---|---|---|---|
| DRZV | D 125 | 1 | 0,141 | 0,088 | 1,6 | Négatif |
| | D 169 | 1 | 0,111 | 0,089 | 1,3 | Négatif |
| | D 170 | 1 | 0,127 | 0,099 | 1,3 | Négatif |
| | D 313 | 1 | 0,178 | 0,102 | 1,8 | Négatif |
| MPE | PBL* | 1 | 0,121 | 0,09 | 1,3 | Négatif |
| Contrôle (AAHL) | R 624 + R 710 | 3 tests | 1,61 | 0,207 | 7,8 | Positif |
| | R 837 U | 3 tests | 0,095 | 0,049 | 1,9 | Négatif |

*:Leucocyte d'un porc malade pendant l'épizootie de PPA

## 2. Recherche d'antigènes NS$_3$ et/ou E$_2$ dans les leucocytes ou rates des bovins

Les 69 échantillons de PBL bovins dont 53 prélevés à l'abattoir frigorifique d'Antananarivo et 16 en provenance de la ferme de KIANJASOA ont tous donné des résultats négatifs au PACE en utilisant les monoclonaux spécifiques du virus BVDV (Tab. V). Il en était de même des 45 prélèvements de rate d'abattoir. L'absence de résultat positif indique qu'aucun échantillon testé ne renferme des antigènes du virus de la maladie des muqueuses.

Les contrôles (BVDV positif et BVDV négatif) fournis par l'AAHL ont tous donné des valeurs prévues par le kit.

Tableau V : Prélèvements bovins : recherche d'antigènes BVDV (NS$_3$ ou/et E$_2$)

| Origine | Echantillons | Nombres | Résultats densité optique (DO) | | Résultats) (S/N |
|---|---|---|---|---|---|
| | | | mAbs R 882 | mAbs R 769 | |
| Abattoir Frigorifique | PBL - Lot 41/98 | 13 | 0.137±0.026 | 0.068±0.010 | Négatif |
| | PBL - Lot 42/98 | 6 | 0.068±0.010 | 0.075±0.010 | Négatif |
| | PBL - Lot 43/98 | 5 | 0.293±0.069 | 0.199±0.071 | Négatif |
| | PBL - Lot 44/98 | 10 | 0.199±0.071 | 0.216±0.042 | Négatif |
| | PBL - Lot 45/98 | 10 | 0.098±0.007 | 0.075±0.008 | Négatif |
| | PBL - Lot 49/98 | 9 | 0.075±0.008 | 0.039±0.006 | Négatif |
| | Broyat de rate | 45 | 0.060±0.013 | 0.038±0.006 | Négatif |
| KIANJASOA | PBL | 16 | 0.056±0.003 | 0.055±0.02 | Négatif |
| AAHL - Australie | *BVDV P++ | 8 tests | 1.105±0.169 | 0.101±0.069 | Positif |
| | **BVDV P+ | 8 tests | 0.429±0.111 | 0.088±0.062 | Positif |
| | ***BVDV négatif | 8 tests | 0.227±0.175 | 0.074±0.039 | Négatif |

\* : Broyat de rate d'un bovin positif en BVDV c'est-à-dire présence d'antigène NS3 ou/et E2 dilué ¼ (AAHL-Australie)

\*\* : Broyat de rate d'un bovin positif en BVDV c'est-à-dire présence d'antigène NS3 ou/et E2 dilué 1/40 (AAHL-Australie)

\*\*\* : Broyat de rate d'un bovin négatif en BVDV c'est-à-dire absence d'antigène NS3 ou/et E2 (AAHL-Australie)

### B. Application du CTB-ELISA pour la recherche d'anticorps anti-pestiviraux

#### 1. Dosage d'anticorps anti-PPC chez les porcs vaccinés, contrôle de qualité du vaccin utilisé

La sérologie des 70 porcs de la ferme de BEVALALA et 10 porcelets de l'IMVAVET, tous vaccinés avec RAMJIVAX ® a montré que la totalité des animaux testés ont réagi positivement : leurs pourcentages d'inhibition en présence de R 1048 varient de 41,4 à 79,3 (cas de la ferme BEVALALA) et de 35,9 à 77,1 (cas de l'IMVAVET). Quant aux témoins non vaccinés (5 porcelets de l'IMVAVET), leurs sérums n'ont donné que des pourcentages d'inhibition allant de 2,5 à 9,1 (Tab. VI), valeurs correspondant à des résultats négatifs.

Tableau VI : Sérologie porcine pour recherche d'anticorps anti-$NS_3$

| Sérum | | CTB-ELISA R 1327 U/ mAbs 1048 | | | | |
|---|---|---|---|---|---|---|
| Origine | Nombre | Densité optique (nm) | | Pourcentage d'inhibition (%) | | Résultats du test |
| | | (a)$DO_{max}$ | (b)$DO_{min}$ | % $Inh_{min}$ | % $Inh_{max}$ | |
| Bev/vacciné | 70 | 1.151 | 0.407 | 41,4 | 79,3 | Positif |
| IMV/vacciné | 10 | 1.260 | 0.450 | 35,9 | 77,1 | Positif |
| IMV/non vacciné | 5 | 1.916 | 1.786 | 2,5 | 9,1 | Négatif |
| AAHL négatif | 1 | 1.965 | | - | | - |
| # 327 | 1 | 0.084 | | 95,7 | | |
| # 330 | 1 | 0.073 | | 96,3 | | Positif |
| # 329 | 1 | 0.079 | | 96 | | |

Pour savoir à quel groupe les anticorps détectés appartiennent, une partie des mêmes sérums a été retestée en présence, respectivement des monoclonaux Tü 24/10 spécifiques du CSFV et des monoclonaux R 882 spécifiques du BVDV (Tab. VII et VIII).

Tableau VII : Sérologie porcine pour recherche d'anticorps anti-$E_2$ spécifique du CSFV

| Sérum | | CTB-ELISA R 1327 U/ mAbs Tü 24/10 | | | | |
|---|---|---|---|---|---|---|
| Origine | Nombre | Densité optique (nm) | | Pourcentage d'inhibition (%) | | Résultats du test |
| | | (a)$DO_{max}$ | (b)$DO_{min}$ | % $Inh_{min}$ | % $Inh_{max}$ | |
| Bev/vacciné | 40 | 0,293 | 0,724 | 35,8 | 74 | Positif |
| IMV/vacciné | 10 | 0,347 | 0,671 | 39,3 | 68,6 | Positif |
| IMV/non vacciné | 5 | 1.068 | 1.015 | 3,4 | 8,2 | Négatif |
| AAHL négatif | 1 | 1.127 | | - | | - |
| # 327 | 1 | 0.084 | | 95,7 | | |
| # 330 | 1 | 0.073 | | 96,3 | | Positif |
| # 329 | 1 | 0.079 | | 96 | | |

Tableau VIII : Sérologie porcine pour recherche d'anticorps anti-$E_2$ spécifique du BVDV

| Sérum | | CTB-ELISA R 1237 U/ mAbs R 882 | | | | |
|---|---|---|---|---|---|---|
| Origine | Nombre | Densité optique (nm) | | Pourcentage d'inhibition (%) | | Résultats du test |
| | | (a)$DO_{max}$ | (b)$DO_{min}$ | % $Inh_{min}$ | % $Inh_{max}$ | |
| Bev/vacciné | 40 | 1.075 | 1.066 | 2,8 | 3,6 | Négatif |
| AAHL négatif | 1 | 1.106 | | - | | - |
| R 947 - C1 | 1 | 0.079 | | 92,9 | | |
| R 947 - C2 | 1 | 0.074 | | 93,3 | | Positif |
| R 947 - C3 | 1 | 0.090 | | 91 | | |

(a) : DOmax : *Densité optique maximale*
(b) : DOmin : *Densité optique minimale*

Ces tests d'identification ont permis d'affirmer que les anticorps vaccinaux renferment non seulement des anticorps anti-p80 (la protéine non structurale $NS_3$) mais aussi des anticorps anti-$E_2$ du virus CSFV car les sérums étudiés inhibent autant R 1048 que Tü 24/10 (Tab. VI et VII). L'absence d'effet inhibiteur des mêmes sérums vis-à-vis de R 882 (mAbs spécifiques du BVDV) est en faveur de l'absence du contaminant BVDV dans le vaccin RAMJIVAX® (Tab. VIII).

L'épreuve virulente effectuée sur les 15 porcelets de l'IMVAVET dont 10 immunisés et 5 témoins avec la souche virulente locale de CSFV, a confirmé la résistance de tous les vaccinés alors que les témoins ont succombé (Tab. IX).

Tableau IX : Test d'efficacité par épreuve virulente

| Porcelets | Résultat | | |
|---|---|---|---|
| | Morts | Vivants | Total |
| Vaccinés | 0 | 10 | 10 |
| Non vaccinés | 5 | 0 | 5 |

Ces résultats d'épreuve comparés à ceux du CTB-ELISA ont montré qu'il existe entre les deux tests une corrélation étroite : tous les porcelets non vaccinés (témoins) ont donné des pourcentages d'inhibition faibles dont les valeurs appartiennent aux négatifs du test ; aucun de ces animaux n'a résisté à l'épreuve virulente. Par contre, ceux qui ont donné des résultats positifs (les vaccinés) ont tous résisté quel que soit son pourcentage d'inhibition (allant de 35,9 à 77,1). Selon ces résultats : les anticorps rencontrés sont spécifiques du CSFV autrement dit le vaccin est immunogène RAMJIVAX® et le test CTB-ELISA peut être utilisé pour remplacer l'épreuve virulente.

2. Recherche des anticorps anti-BVDV (anti-$E_2$)

Pour la recherche des anticorps anti-$E_2$, nous avons utilisé l'antigène BVDV R 1237 (un broyat de rate de bovin positif en BVDV - AAHL /Australie) et le mAbs R 882, spécifique du BVDV.

Tous les sérums analysés, qu'ils soient de veau (KIANJASOA, BEVALALA) ou des bovins adultes (les pools de sérums récoltés à l'abattoir frigorifique d'Antananarivo) ont donné des résultats négatifs au CTB-ELISA. Leurs pourcentages d'inhibition (%INH) sont tous inférieurs à 15%, seuil de positivité du test (Tab. X).

Les valeurs obtenues avec les réactifs de contrôle se sont retrouvées dans les limites attendues, indiquant que les réactions ont été effectuées dans les conditions optimales : le sérum de contrôle positif (R957 – AAHL) est positif quelle que soit sa dilution (jusqu'au 1/500), les pourcentages d'inhibition diminuent graduellement en fonction de la dilution ; les sérums normaux de référence (ABS-AAHL, NHS-AAHL) sont restés négatifs.

Tableau X : <u>Sérologie bovine pour recherche d'anticorps anti-$E_2$ spécifique du BVDV</u>

| Sérum | | CTB-ELISA R 1237-U/ mAbs R 882 | | | | |
|---|---|---|---|---|---|---|
| Origine | Nombre | Densité optique (nm) | | Pourcentage d'inhibition (%) | | Résultats du test |
| | | (a)DO $_{max}$ | (b)DO $_{min}$ | % Inh $_{min}$ | % Inh $_{max}$ | |
| KIANJASOA | 18 | 1.842 | 1.526 | + 11,57 | 7,87 | Négatif |
| BEVALALA | 8 | 1.792 | 1.550 | + 08,54 | 6,12 | Négatif |
| ABATTOIR | 13 | 1.795 | 1.526 | + 08,72 | 7,57 | Négatif |
| ABS* | 1 | 1.651 | | référence | | - |
| NHS** | 1 | 1.910 | | + 15.7*** | | Négatif |
| R 947 - C1 | 1 | 0.257 | | 84,43 | | Positif fort |
| R 947 - C2 | 1 | 0.594 | | 64,02 | | Positif moyen |
| R 947 - C3 | 1 | 1.04 | | 37,01 | | Positif faible |

\* : ABS : Sérum de bovin exempt de BVDV utilisé comme contrôle négatif
\*\*\* : Sérum Normal de cheval ou NHS
\*\*\* : les signes positifs indiquent que la valeur de la DO est supérieure au contrôle négatif, donc tous les résultats précédés de ce signe sont considérés comme zéro pour cent d'inhibition
(a) : DOmin : Densité optique minimale
(b) : DOmax : Densité optique maximale

## IV. **DISCUSSION**

Les prélèvements en provenance des différents élevages victimes de cette épidémie porcine malgache de 1998 ont été testés au PACE et ont donné des résultats négatifs dans la majeure partie (15 prélèvements sur 19). Le PACE étant un test hautement spécifique (Sp environ 100%) [7], la probabilité pour que ces négatifs du test soient des vrais négatifs, est très élevée (VPN = 99,7%) [7]. Pendant ce temps, le diagnostic de laboratoire du Centre National d'Etudes Vétérinaires et Alimentaires ou CNEVA (Maisons Alfort - France) réalisé sur des échantillons provenant des élevages encadrés par la Maison du Petit Elevage (MPE) a été formel en précisant que l'épizootie meurtrière touchant le cheptel porcin de Madagascar en décembre 1998 était due à la PPA (communication de la Direction des Services Vétérinaires 1999), une virose porcine qui n'a aucune antigénicité croisée avec la PPC. Ce résultat est en parfaite concordance avec celui dont nous avons pu avoir au PACE, situation en faveur du renforcement de la qualité de ce test. Quant aux résultats positifs donnés par le test (4/19), vu sa grande sensibilité (Se environ 99 %) et sa forte valeur prédictive positive (VPP environ 100%) [7] : la chance pour que ces 4 cas soient des vrais positifs PPC est très élevée. Cette présence de victimes du CSFV seul ou en co-infection avec la PPA en pleine épizootie de cette dernière traduit une situation épidémiologique parfaitement normale pour le cas de Madagascar où la PPC reste endémique depuis son introduction en 1965 (Maminiaina 1999).

La recherche du virus BVDV s'est également avérée négative en appliquant les deux tests, le PACE pour la détection d'antigènes ($NS_3$ et/ou $E_2$) et le CTB-ELISA pour les anticorps correspondants (anti-$E_2$). Des investigations ont été réalisées auparavant par une équipe de chercheurs de l'IEMVT [2] et la conclusion a été en faveur de l'absence de cette maladie dans la grande Ile. A défaut d'autres résultats scientifiques confirmant le contraire, et tenant compte des origines variées des bovins abattus à l'abattoir frigorifique d'Antananarivo, nous pouvons admettre l'absence du BVDV à Madagascar.

Il en est de même de l'application des deux réactions immuno-enzymatiques dans le contrôle de qualité du vaccin local anti-PPC RAMJIVAX® : les résultats obtenus concordent parfaitement avec ceux des épreuves virulentes. Cette concordance est tout à fait logique dans la mesure où les pourcentages d'inhibition du CTB-ELISA, qui intéressent uniquement la protéine non structurale p80 reflètent théoriquement ceux des tests de séroneutralisation, vu la valeur très élevée de la corrélation entre le taux des anticorps anti-$NS_3$ et celui des anticorps neutralisants (r = 0,98) [communication personnelle de Shannon].

## V. **CONCLUSION**

Le but de cette étude, menée dans le cadre d'un projet de coopération technique avec l'AIEA est de voir comment les deux systèmes ELISA antigènes et anticorps se comportent dans les conditions de Madagascar. Il en ressort les qualités générales du test enzymo-immunologique : rapidité, sensibilité et spécificité. Le CTB-ELISA dont la corrélation avec la réaction de neutralisation virale est très élevée (r = 0,98) apporte à notre laboratoire des améliorations importantes par rapport aux techniques traditionnelles d'épreuves virulentes et de séroneutralisation. Il met également en évidence l'efficacité des vaccinations anti- PPC pratiquées avec RAMJIVAX®. Les résultats de l'utilisation des deux types d'ELISA viennent par ailleurs confirmer ceux de Blancou et ces collaborateurs à propos de l'absence du virus de la maladie des muqueuses des bovins à Madagascar.

Les deux tests ELISA PACE et CTB-ELISA, sont des outils de diagnostic rapide, facile à mettre en œuvre, adaptés à l'évaluation de la réponse immunitaire post-vaccination et à des enquêtes épidémiologiques à grande échelle sur les infections causées par les Pestivirus bovins et porcins.

**Remerciements**

**Références bibliographiques**